\documentclass{ws-procs9x6}            
\begin{document}
\title{$\Lambda^*$ matter and its stability 
}

\author{J. Hrt\'{a}nkov\'{a}$^{a,*}$, M. Sch\"{a}fer$^{a,b}$, and J. Mare\v{s}$^{a}$}

\address{$^a$ Nuclear Physics Institute CAS, 250 68 Rez,Czech Republic\\
$^*$E-mail: hrtankova@ujf.cas.cz\\
$^b$ Faculty of Nuclear Sciences and Physical Engineering, Czech 
Technical University in Prague, 115 19 Prague 1, Czech Republic}

\author{A. Gal, E. Friedman, and N. Barnea}

\address{Racah Institute of Physics, The Hebrew University, 91904 Jerusalem, Israel}

\begin{abstract}
We performed calculations of nuclear systems composed solely of $\Lambda^*$ hyperons, aiming at exploring the possibility of existence of absolutely stable $\Lambda^*$ matter. We considered $\Lambda^*$ interaction strengths compatible with the 
$\Lambda^*\Lambda^*$ binding energy $B_{\Lambda^*\Lambda^*}$ given by the 
$\bar{K}N$ interaction model by Yamazaki and Akaishi~\cite{YA07}. We found that the binding energy per $\Lambda^*$ saturates at values well below 100 MeV for mass number 
$A\geq120$. The $\Lambda^*$ matter is thus highly unstable against strong interaction decay. 
\end{abstract}


\bodymatter

\section{Introduction}
This contribution concerns our recent study of $\Lambda^*$ nuclei~\cite{HBFGMS18}, 
which was stirred up by a conjecture about absolutely stable charge-neutral 
baryonic matter composed solely of $\Lambda(1405)$ ($\Lambda^*$) hyperons\cite{AY17}.\\
We calculated $\Lambda^*$ few-body systems within the Stochastic Variational Method (SVM)~\cite{VS98}, as well as $\Lambda^*$ many-body systems within the Relativistic Mean Field (RMF) approach~\cite{SW86}. 
The meson-exchange $\Lambda^*$ potentials applied in our work were fitted to reproduce the $\Lambda^*\Lambda^*$ binding energy 
$B_{\Lambda^*\Lambda^*}=40$~MeV, given by the phenomenological ${\bar K}N$ interaction model~\cite{YA07}. We recall that the ${\bar K}N$ potentials 
used by Akaishi and Yamazaki~\cite{YA07, AY17}, fitted for $I=0$ to the mass and width of the $\Lambda(1405)$ resonance, fail to reproduce $K^-$ single-nucleon absorption fractions deduced from $K^-$ capture bubble chamber experiments\cite{FG17}. 
Nevertheless, we employed these very strong potentials in order to demonstrate 
that while solving the $A$-body Schr\"{o}dinger equation for purely attractive 
$\Lambda^*\Lambda^*$ interactions will inevitable lead to collapse, with the binding energy per particle diverging as $A$ increases, this scenario promoted 
in ref.\cite{AY17} is unlikely in standard many-body approaches.\\ 
In the following sections, we discuss only briefly our main results; more details can be found in ref.\cite{HBFGMS18}.

\section{$\Lambda^*$ Few-Body Systems}
We started our study of $\Lambda^*$ nuclei by calculations of few-body systems 
within the Stochastic Variational Method~\cite{VS98} for the meson-exchange potentials of the Dover-Gal form~\cite{DG84}:\\ 
 \begin{eqnarray}
V_{\Lambda^*\Lambda^*}(r) =& g_{\omega \Lambda^*}^2 \,(1+{\frac{1}{8}\frac{m_{\omega}^2}{M_{\Lambda^*}^2}})\,
Y_{\omega}(r) - g_{\sigma \Lambda^*}^2 \,(1-{\frac{1}{8}\frac{m_{\sigma}^2}{M_{\Lambda^*}^2}})\,
Y_{\sigma}(r)\\& + g_{\omega \Lambda^*}^2 {\frac{1}{6}\frac{m_{\omega}^2}{M_{\Lambda^*}^2}}\,Y_{\omega}(r) {(\vec{\sigma}_1 \cdot \vec{\sigma}_2)}~, \nonumber 
\label{eq:DG}
\end{eqnarray} 
    or the Machleidt form~\cite{Mach88}: 
\begin{eqnarray} 
V_{\Lambda^*\Lambda^*}(r) =& g_{\omega \Lambda^*}^2 (1+{\frac{1}{2}\frac{m_{\omega}^2}{M_{\Lambda^*}^2}})\,Y_{\omega}(r) - g_{\sigma \Lambda^*}^2 \, 
(1-{\frac{1}{4}\frac{m_{\sigma}^2}{M_{\Lambda^*}^2}})\, Y_{\sigma}(r) \\& + g_{\omega \Lambda^*}^2 {\frac{1}{6}\frac{m_{\omega}^2}{M_{\Lambda^*}^2} }\,Y_{\omega}(r) {(\vec{\sigma}_1 \cdot \vec{\sigma}_2)}~, 
\nonumber
\label{eq:Mach}
\end{eqnarray} 
where $M_{\Lambda^*}=1405$~MeV, $m_i$ are the meson masses, $g_{i \Lambda^*} = \alpha_i g_{i N}$ are the corresponding coupling constants with $g_{i N}$ taken from the HS model~\cite{HS81}, and 
$Y_{i=\sigma, \omega}(r)=\exp(-m_i r)/(4\pi r)$. 
In the above expressions, the mass correction factors 
($\sim m^2_{i}/M^2_{\Lambda^*}$) as well as the spin-spin interaction terms ($\sim (\vec{\sigma}_1 \cdot \vec{\sigma}_2$)) are included.

In the calculations we fit either the value of $\alpha_{\sigma}$  and kept $\alpha_{\omega}$ fixed to 1 or vice versa in order to get the binding energy of the $\Lambda^*\Lambda^*$ system $B_{\Lambda^*\Lambda^*}=40$~MeV.
We present here only selected results for $\alpha_{\sigma}\neq 1$. 

\begin{figure}
\begin{center}
\includegraphics[width=0.45\textwidth]{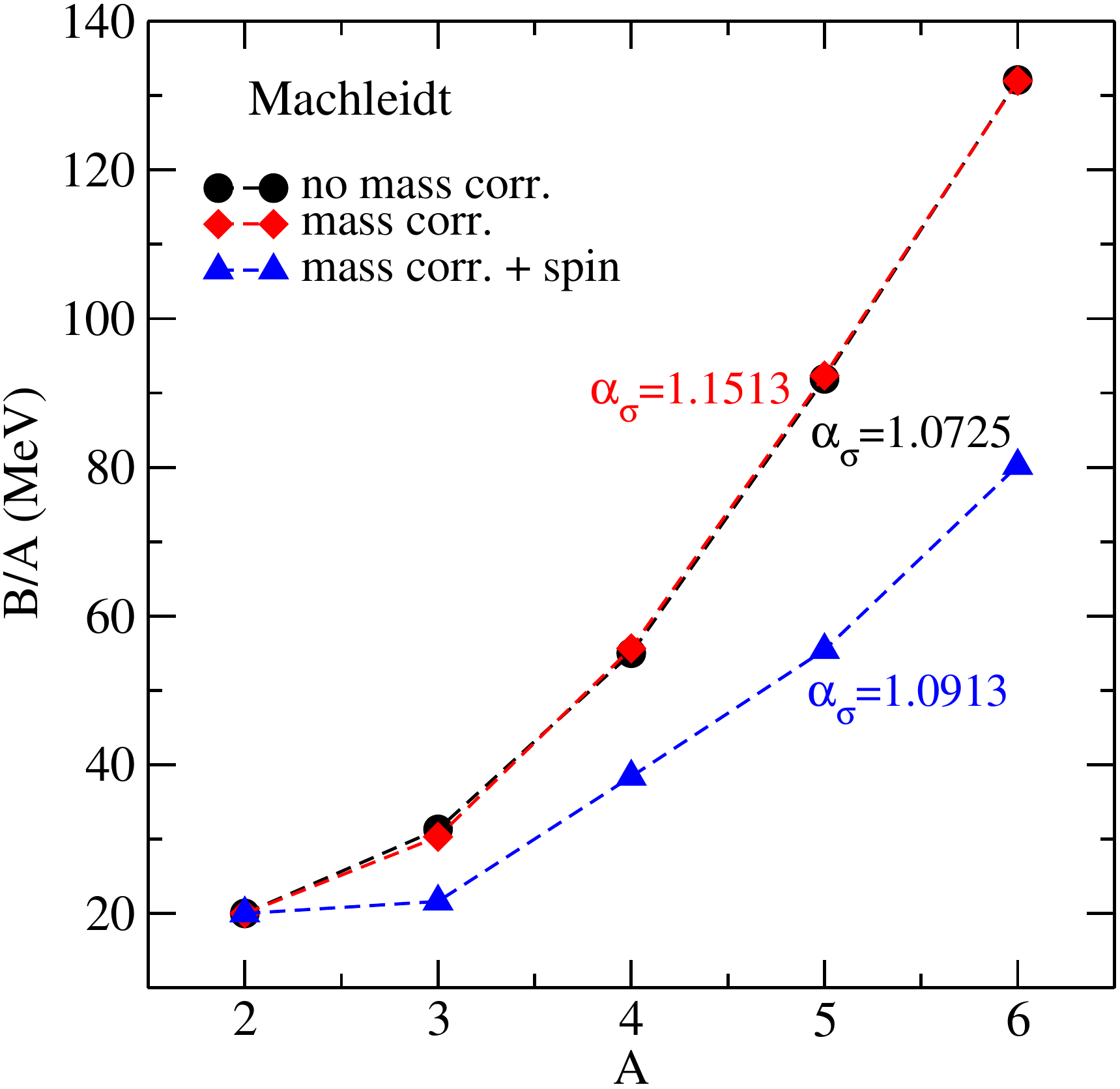}\hspace{5pt}
\includegraphics[width=0.45\textwidth]{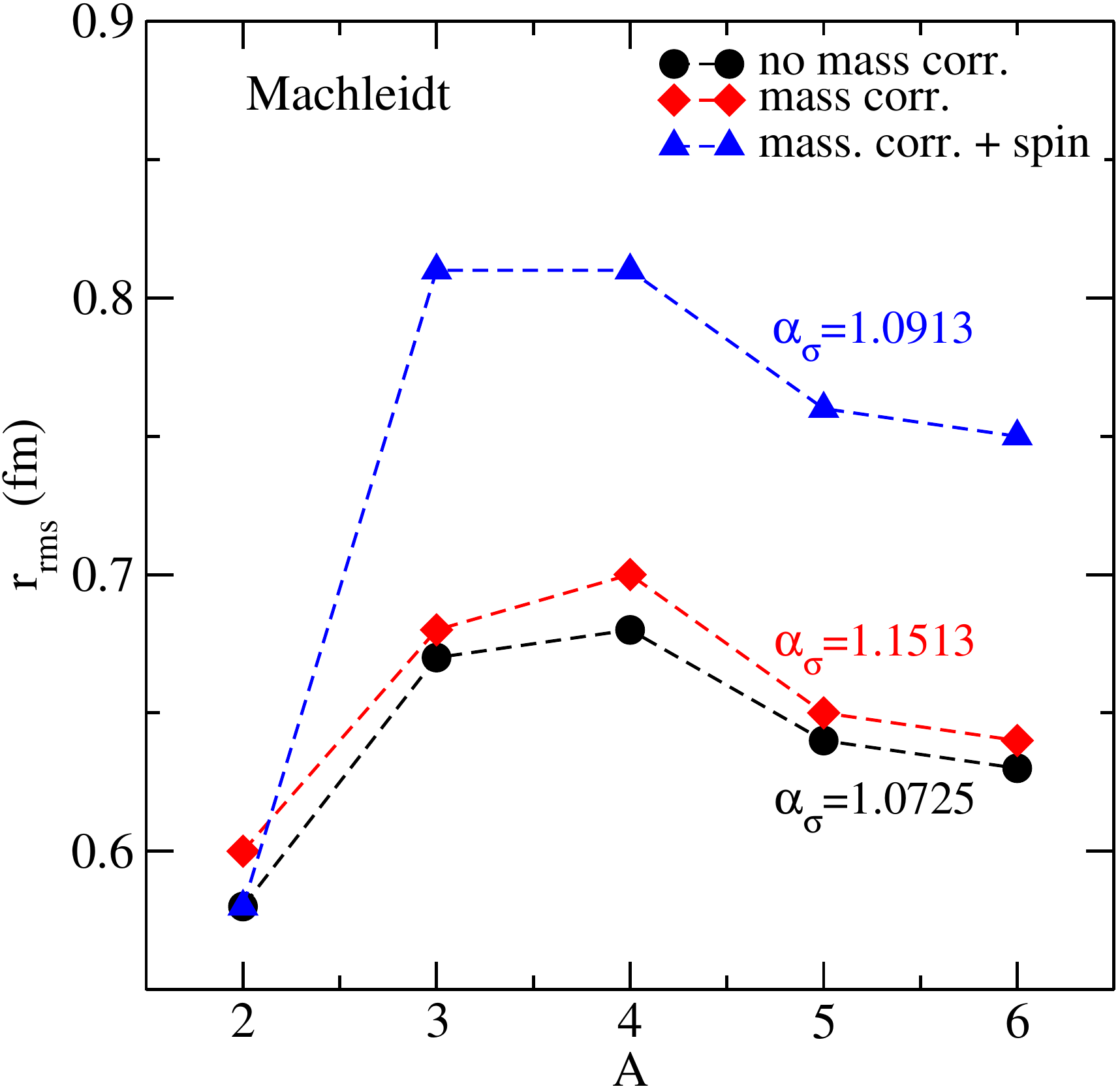}
\end{center}
\caption{Binding energy of $\Lambda^{\ast}$ nuclei per particle, $B/A$ (left panel) and rms radius (right panel) of few-body $\Lambda^*$ systems as a function of mass number $A$, calculated using the Machleidt potential with and without mass corrections, as well as including spin-spin interaction.}
\label{SVM}
\end{figure}

In Fig.~\ref{SVM}, left panel, we show the binding energy per $\Lambda^*$, B/A, as a function of mass number in few-body $\Lambda^*$ nuclei, calculated within the SVM approach for the Machleidt potential (1). When the spin-spin interaction is omitted, the binding energy per particle is rapidly increasing with A, reaching $B/A \approx 130$~MeV for A=6. The mass corrections have almost no effect on the calculated values of B/A. On the other hand, when the spin-spin interaction is taken into account, the increase of B/A is considerably less steep. 
The corresponding rms radius of the considered $\Lambda^*$ nuclei is presented in the right panel. The rms radius is extremely small, hardly exceeding the value 0.8~fm even if the spin-spin interaction is included. 

\section{$\Lambda^*$ Many-Body Systems}
As the next step, we explored many-body systems composed solely of $\Lambda^*$ hyperons within the RMF framework \cite{SW86}, where the interaction among $\Lambda^*$'s 
is mediated by the exchange of the scalar $\sigma$ and vector $\omega$ meson fields. The underlying Lagrangian density is of the form
\begin{equation}  
\mathcal L = \bar{\Lambda}^* \left[\,{\rm i}\gamma^\mu D_\mu-(M_{\Lambda^*}-g_{\sigma \Lambda^*} \sigma) \right] \Lambda^* + (\sigma, \omega_\mu\,\textrm{free-field terms})~,
\label{eq:Lag} 
\end{equation}
where $D_\mu=\partial_\mu+{\rm i}\,g_{\omega \Lambda^*}\,\omega_\mu$. 
It is to be noted that the isovector-vector $\vec{\rho}$ and Coulomb fields were not taken into account since the $\Lambda^*$ is a neutral $I=0$ baryon.
First calculations were perfomed using the linear HS model~\cite{HS81} with the coupling constants scaled by $\alpha_{\i}$, $g_{i \Lambda^*} = \alpha_i g_{i N}$, determined by fitting $B_{\Lambda^*\Lambda^*}$ (see previous section). 
For comparison, we performed also calculations using the nonlinear NL-SH  model~\cite{NLSH}. The corresponding scaling parameter $\alpha_{\sigma}$ was fitted to yield the binding energy of the $8\Lambda^*$ system calculated within the HS model. 
We explored $\Lambda^*$ nuclei with closed shells and solved self-consistently the coupled system of the Klein-Gordon equations for meson fields and the Dirac equation for $\Lambda^{\ast}$. 

\begin{figure}[t]
\begin{center}
\includegraphics[width=0.45\textwidth]{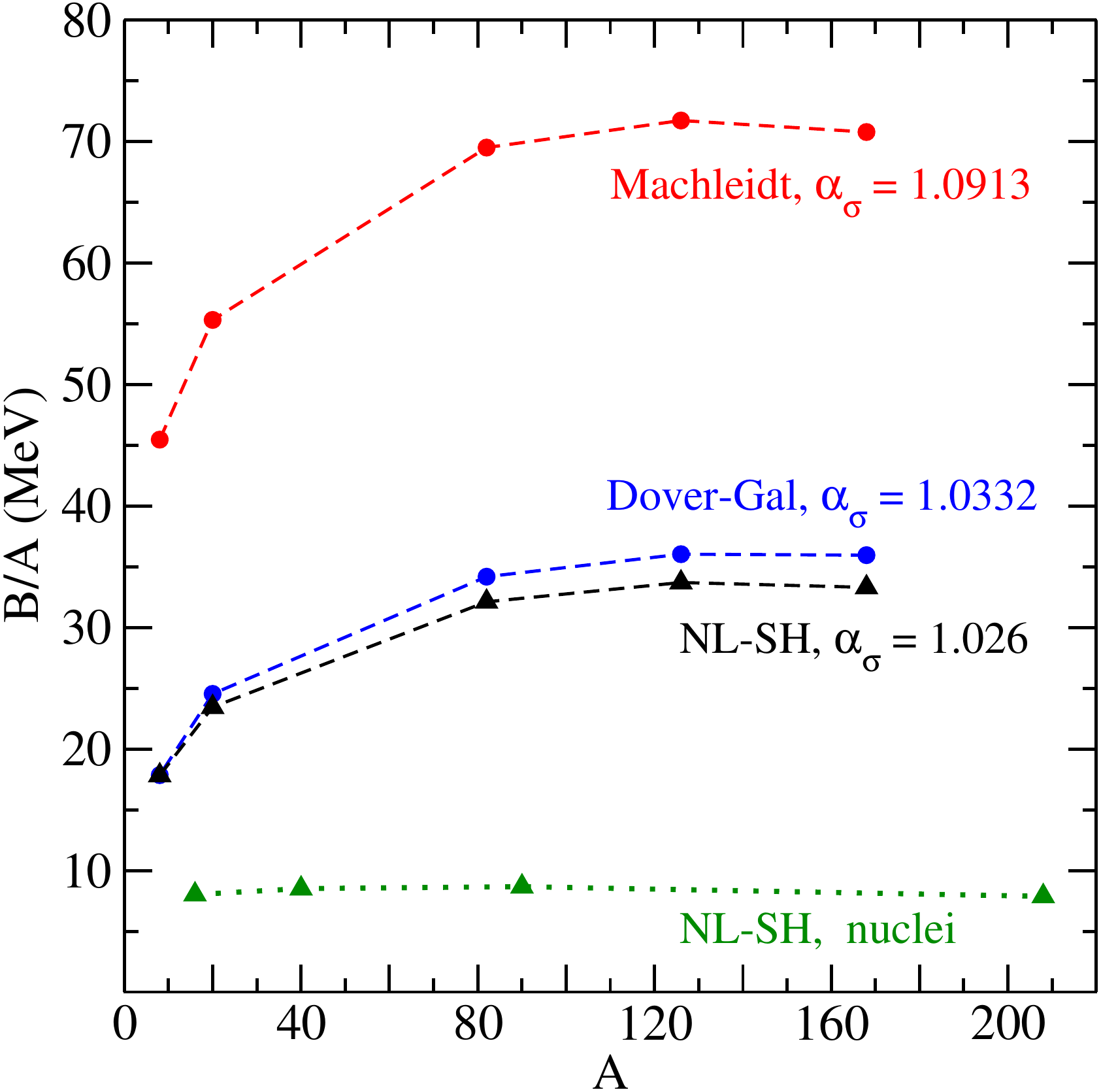}\hspace{5pt}
\includegraphics[width=0.46\textwidth]{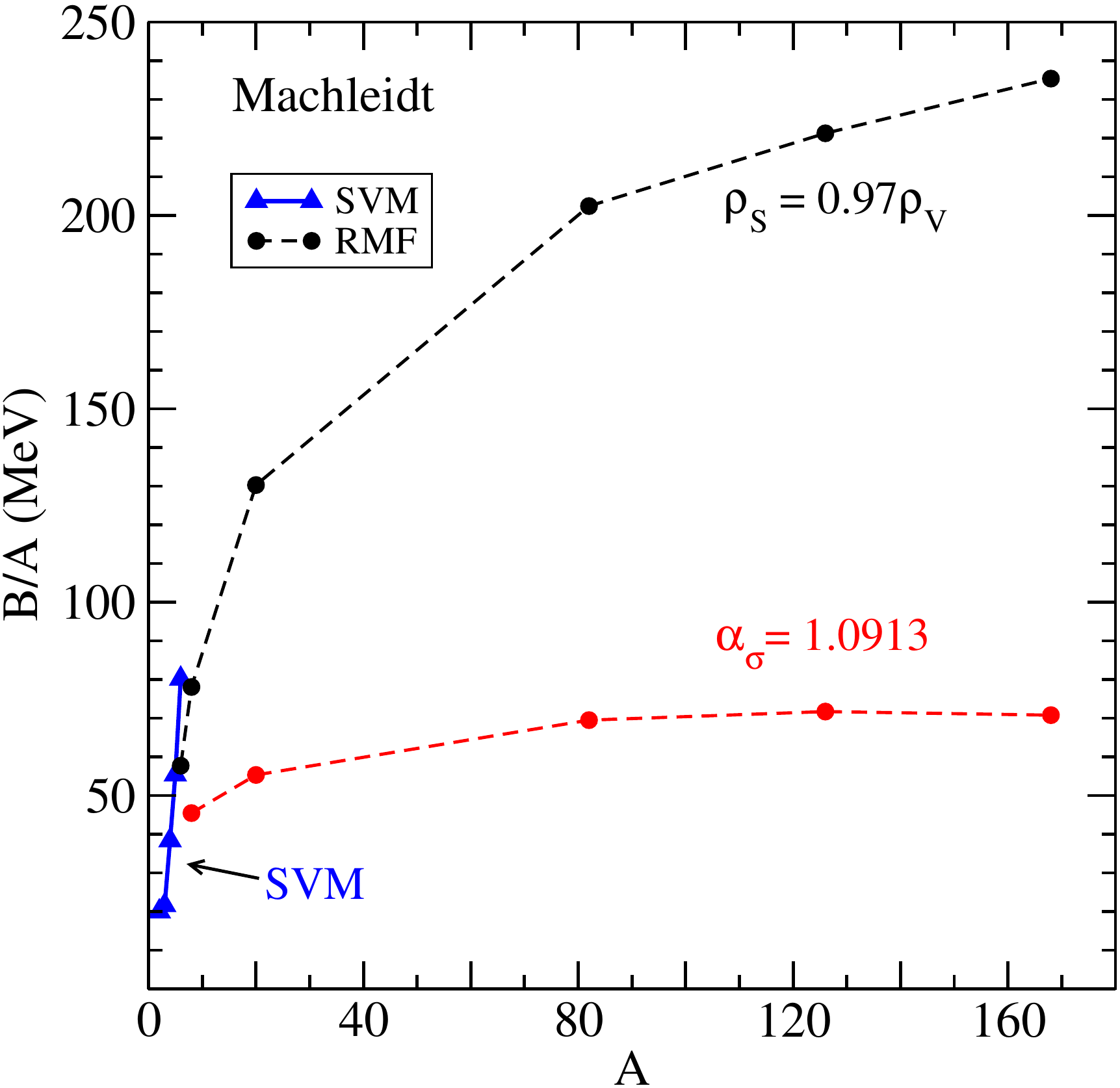}
\end{center}
\caption{Left panel: Binding energy of $\Lambda^{\ast}$ nuclei per particle, $B/A$ as a function of mass number $A$, calculated within the HS 
and NL-SH models; $B/A$ in atomic nuclei (`nuclei') is shown for comparison. Right panel: Comparison of $B/A$ 
calculated in $\Lambda^*$ nuclei within the HS model for the Machleidt potential (red line) with a similar calculation using $\rho_{\rm s}=0.97\rho_{\rm v}$ (black line); $B/A$ in few body systems, calculated within the SVM is shown for comparison. See text for details.
}
\label{RMF}
\end{figure}

The results of our RMF calculations are summarized in Fig.~\ref{RMF}. 
In the left panel, the binding energy per particle, $B/A$, is plotted as a function of mass number $A$, calculated within the RMF HS model with the properly rescaled $\sigma$ meson coupling constant corresponding to the 
$\Lambda^*$ potentials (1) and (2). For comparison, $B/A$ calculated within the RMF NL-SH model in $\Lambda^*$ nuclei as well as in ordinary nuclei is shown as well. The binding energy per $\Lambda^*$ saturates with the number of constituents for $A\geq120$ in all versions considered and reaches tens of MeV depending on the potential used. Calculations with the rescaled $\omega$ coupling constant yield similar saturation curves for B/A in $\Lambda^*$ nuclei. 

The observed saturation originates from the Lorentz covariance which introduces two types of baryon densities --- the scalar density $\rho_{\rm s}$ associated with the attractive $\sigma$ field and the vector (baryon) density $\rho_{\rm v}$ associated with the repulsive $\omega$ field. In dense matter, the scalar density decreases with respect to the vector density since $\rho_{\rm s} \sim {M^*}/{E^*} \rho_{\rm v} \quad \textrm{where} \quad \frac{M^*}{E^*} < 1~,$  and $M^*= M-g_{\sigma B} \langle \sigma \rangle$ is baryon effective mass. As a consequence, the attraction from the scalar field is reduced considerably at higher densities. This is illustrated in Fig.~\ref{RMF} (right panel), where we present the RMF calculation of $B/A$ in $\Lambda^*$ nuclei, in which we replaced the scalar density $\rho_{\rm_s}$ by a density equal to $0.97 \rho_{\rm v}$ (this corresponds to $\rho_{\rm s}/\rho_{\rm v}$ in $^{16}$O). The binding energy per $\Lambda^*$ (denoted '$\rho_{\rm s }=0.97\rho_{\rm v}$') is rapidly increasing in this case, similar to the SVM calculations (also shown for comparison), and does not seem to saturate within the explored mass range, unlike B/A evaluated using the  'dynamical' scalar density $\rho_{\rm s}$ (denoted '$\alpha_{\sigma} = 1.0913$'). It is to be noted that the central density of calculated $\Lambda^*$ nuclei saturates as a function of $A$ as well, reaching about twice nuclear matter density.   

Finally, we introduced the $\Lambda^*$ absorption and explored how the $\Lambda^*$ decay width changes in the medium. We considered the two-body decay $\Lambda^*\Lambda^* \rightarrow \Lambda \Lambda$ in the $1s$ state, described by the imaginary part of an optical potential in a '$t\rho$' form with the amplitude fitted to assumed width 
$\Gamma_{\Lambda^*\Lambda^*} = 100$~MeV at threshold, taking into account 
phase space suppression. We found that the conversion widths, despite being  suppressed to some extent in the $\Lambda^*$ nuclei (by ~28\% in A=8 systems and by less than 1\% in A=168 systems), remain considerable and the $\Lambda^*\Lambda^*$ pairs will thus inevitably decay.

\section{Summary}

We performed calculations of $\Lambda^*$ nuclei with various $\Lambda^*$ interaction strengths compatible with the value $B_{\Lambda^* \Lambda^*}=40$~MeV of the YA model~\cite{YA07} in order to demonstrate that the $\Lambda^*$ stable-matter scenario~\cite{AY17} is not supported by standard many-body approaches. We found that the binding energy per $\Lambda^*$ in many-body systems saturates in all cases for $A\geq 120$ at values far below $\approx 290$~MeV, which is the energy required to reduce the $\Lambda(1405)$ mass in the medium below the mass of the lightest hyperon $\Lambda(1116)$. The $\Lambda^*$ matter is thus highly unstable against strong interaction decay.

\section*{Acknowledgment}
This work was partly supported by the Czech Science Foundation GACR grant 19-19640S.

\end{document}